\documentclass[12pt,showpacs,preprintnumbers,superscriptaddress,amsmath,amssymb,nofootinbib]{revtex4}
\usepackage{graphicx}
\usepackage{dcolumn}
\usepackage{bm}
\usepackage{amssymb}
\usepackage{amsmath}
\usepackage{epsfig}    
\usepackage{color}
\usepackage{hhline}
\def\be{\begin{equation}}
\def\ee{\end{equation}}
\newcommand{\bea}{\begin{eqnarray}}
\newcommand{\eea}{\end{eqnarray}}
\newcommand{\nn}{\nonumber}

\numberwithin{equation}{section}

\begin{document}


\title{ Radiative Seesaw Model with  Degenerate Majorana Dark Matter }

\preprint{KIAS-P16024}
\author{Takaaki Nomura}
\email{nomura@kias.re.kr}
\affiliation{School of Physics, KIAS, Seoul 130-722, Korea}

\author{Hiroshi Okada}
\email{macokada3hiroshi@gmail.com}
\affiliation{Physics Division, National Center for Theoretical Sciences, Hsinchu, Taiwan 300}

\author{Yuta Orikasa}
\email{orikasa@kias.re.kr}
\affiliation{School of Physics, KIAS, Seoul 130-722, Korea}
\affiliation{Department of Physics and Astronomy, Seoul National University, Seoul 151-742, Korea}

\date{\today}

\begin{abstract}
We study a three loop induced  neutrino mass model with exotic vector-like isospin doublet leptons which contain a dark matter candidate.
Then we explore lepton flavor violations, and dark matter physics in co-annihilation system. 
In this paper the nearly degenerate Majorana fermion dark matter can naturally be achieved at the two loop level, while
the mass splitting can be larger than ${\cal O}$(200) keV which is required from the constraint of the direct detection search with spin independent inelastic scattering through $Z$ boson portal.
As a result a monochromatic  photon excess, its threshold energy is greater than ${\cal O}$(200) keV, is predicted in our model that could be measured through indirect detection experiments such as INTEGRAL.
\end{abstract}
\maketitle
\newpage

\section{Introduction}
If the active neutrino could be related to new physics beyond the standard model (SM) such as dark matter (DM) candidate within TeV scale,
it could be verifiable models through various experiments in the near future. In this sense, radiative seesaw models 
could be one of  the natural realization of the tiny neutrino mass at TeV energy scale involving DM.
Along this line of ideas,  a vast paper has been arisen in Refs.~\cite{Zee, Cheng-Li, zee-babu, Krauss:2002px, Ma:2006km, Aoki:2008av, Gustafsson:2012vj, Hambye:2006zn, Gu:2007ug, Sahu:2008aw, Gu:2008zf, Babu:2002uu, AristizabalSierra:2006ri, AristizabalSierra:2006gb,
Nebot:2007bc, Bouchand:2012dx, Kajiyama:2013sza,McDonald:2013hsa, Ma:2014cfa, Schmidt:2014zoa, Herrero-Garcia:2014hfa,
Ahriche:2014xra,Long1, Long2, Aoki:2010ib, Kanemura:2011vm, Lindner:2011it,
Kanemura:2011jj, Aoki:2011he, Kanemura:2011mw, Schmidt:2012yg, Kanemura:2012rj, Farzan:2012sa, Kumericki:2012bf, Kumericki:2012bh, Ma:2012if, Gil:2012ya, Okada:2012np, Hehn:2012kz, Baek:2012ub, Dev:2012sg, Kajiyama:2012xg, Kohda:2012sr, Aoki:2013gzs, Kajiyama:2013zla, Kajiyama:2013rla, Kanemura:2013qva,Law:2013saa, Dasgupta:2013cwa, Baek:2013fsa, Okada:2014vla, Ahriche:2014cda, Ahriche:2014oda,Chen:2014ska,
Kanemura:2014rpa, Okada:2014oda, Fraser:2014yha, Okada:2014qsa, Hatanaka:2014tba, Baek:2015mna, Jin:2015cla,
Culjak:2015qja, Okada:2015nga, Geng:2015sza, Okada:2015bxa, Geng:2015coa, Ahriche:2015wha, Restrepo:2015ura, Kashiwase:2015pra, Nishiwaki:2015iqa, Wang:2015saa, Okada:2015hia, Ahriche:2015loa, Ahn:2012cg, Ma:2012ez, Kajiyama:2013lja, Hernandez:2013dta, Ma:2014eka, Aoki:2014cja, Ma:2014yka, Ma:2015pma, Ma:2013mga,
radlepton1, radlepton2, Okada:2014nsa, Brdar:2013iea, Okada:2015nca, 
Okada:2015kkj, Fraser:2015mhb, Fraser:2015zed, Adhikari:2015woo, Kanemura:2015cca, Bonnet:2012kz,Sierra:2014rxa, Davoudiasl:2014pya, Lindner:2014oea,Okada:2014nea, MarchRussell:2009aq, King:2014uha, Mambrini:2015sia, Boucenna:2014zba, Ahriche:2016acx, Okada:2015vwh, Ahriche:2016rgf, Nomura:2016run}. 

In this paper,  a three loop induced  neutrino mass model is studied where
we analyze lepton flavor violations, and dark matter physics in co-annihilation system. 
{To generate active neutrino mass, we introduce extra SU(2)$_L$ doublet leptons whose neutral component can be a DM candidate. 
For each doublet, we have two neutral Majorana fermions with degenerate mass at the tree level since the extra leptons are introduced as vector-like, 
which is highly constrained by direct detection search of DM due to $Z$ boson exchanging process.
The splitting of neutral fermion mass degeneracy} can naturally be achieved at the two loop level by introducing extra isospin singlet charged scalar fields. 
As a result, a monochromatic photon excess could be measured through indirect detection experiments such as INTEGRAL~\cite{Jean:2003ci} at 511 keV photon emission energy where its threshold energy is greater than ${\cal O}$(200) keV~\cite{TuckerSmith:2001hy}. 
Here {the lower limit of the energy} comes from the constraint of the direct detection search with spin independent inelastic scattering through $Z$ boson portal. 
Also we discuss the way to increase the mass difference between these degenerated masses at the two loop level through Zee-Babu type diagram. Since this mass is proportional to the charged lepton masses, the natural scale is the active neutrino masses ${\cal O}(0.1-1)$ eV, which is ruled out by the direct detection searches. 
To evade the problem, we introduce additional isospin singlet charged bosons with the same quantum number as the original field contents. 
Then the degenerated mass scale can achieve the order ${\cal O}$(200) keV. However, the muon anomalous magnetic moment is also negatively enhanced by this effect. Here we also discuss the way to evade this constraint.

This paper is organized as follows.
In Sec.~II, we show our model building including Yukawa and Higgs sector. In Sec.~III we discuss phenomenology of the model such that mass formulations, and analysis of the DM physics.
We conclude and discuss in Sec.~IV.

\section{The Model}


\begin{table}[thbp]
\centering {\fontsize{10}{12}
\begin{tabular}{|c||c|c|c|}
\hline Fermion & $L_L$ & $ e_{R} $ & $L'_{}$  
  \\\hhline{|=#=|=|=|$}
$(SU(2)_L, U(1)_{\rm Y})$ & $(\bm{2},-1/2)$ & $(\bm{1}, -1)$ & $(\bm{2},-1/2)$   
\\\hline
$Z_2$ & $+$ & $+$ &  $-$  \\\hline
\end{tabular}%
} \caption{Lepton sector, where the three  flavor index for each field $L_{L}$, $e_R$ and $L'_{}$ is abbreviated.} 
\label{tab:1}
\end{table}

\begin{table}[thbp]
\centering {\fontsize{10}{12}
\begin{tabular}{|c||c|c|c|c|}\hline 
Boson  & $\Phi$      & $h^+_i$   & $k^{++}_i$ & $S$   
 \\\hhline{|=#=|=|=|=|}
$(SU(2)_L,U(1)_{\rm Y})$ & $(\bm{2},1/2)$     & $(\bm{1},1)$   & $(\bm{1},2)$ & $(\bm{1},0)$    \\\hline
$Z_2$ & $+$ & $-$ &  $+$ & $-$ \\\hline
\end{tabular}%
} 
\caption{Boson sector: i runs 1 to N. }
\label{tab:2}
\end{table}

In this section, we explain our model .
For the fermion sector, we introduce {$SU(2)_{L}$} doublet exotic leptons ($L'$) with three flavors to the SM leptons ($L_L$,$e_R$), as we summarize in Tab.~\ref{tab:1}.
For the boson sector, we introduce a {$SU(2)_{L}$} singlet inert scalar ($S$),
{$SU(2)_{L}$ isospin singlet singly charged scalars ($h^{\pm}_i$)and doubly-charged scalars ($k^{\pm\pm}_i$)} to the SM Higgs boson ($\Phi$), as we summarize in Tab.~\ref{tab:2}.
$Z_2$ symmetry plays an role in our radiative neutrino model at the three loop, and assures the stability of the dark matter candidates $L'$ and/or $S$.  
$\Phi$ develops a VEV (denoted by $v/\sqrt2$) after breaking the $SU(2)_L$ symmetry.

The renormalizable Lagrangian for Yukawa sector and scalar potential under these assignments
are relevantly given by
\begin{align}
-\mathcal{L}_{Y}
=&
(y_{\ell})_{ij} \bar L_{L_i} \Phi_L e_{R_j} 
+(f_R)_{ij} \bar L_{L_i} L'_{R_j} S 
+{ \sum_{a=1}^N \left((f_L)_{ij}^a \bar L'_{L_i} i\tau_2 L_{L_j}^C h^-_a
+  g_{ij}^a \bar e^C_i e_j k^{++}_a\right) }
+ (M_{E})_{i} \bar L'_{L_i} L'_{L_i} 
+{\rm c.c.},
 \label{Lag:Yukawa}\\ 
\mathcal{V}
=&
 -m^2_{\Phi} |\Phi|^2  - {\sum_{a=1}^N \left(m^2_{h_a} |h^+_a|^2  + m^2_{k_a} |k^{++}_a|^2\right)} - m^2_{S} (S^2+{\rm c.c.}) - m'^2_{S} |S|^2
 -{\sum_{a,b,c=1}^N\mu_{abc}(h^-_a h^-_b k^{++}_c+{\rm c.c.})}
 \nn\\
&
+\lambda_{\Phi}|\Phi|^4 + {\sum_{a,b,c,d=1}^N \left( \lambda_{h_{abcd}}h^+_a h^+_b h^-_c h^-_d 
+ \lambda_{k_{abcd}}k^{++}_a k^{++}_b k^{--}_c k^{--}_d\right) +\lambda_{{h_{ab}k_{cd}}}h^+_a h^-_b k^{++}_c k^{--}_d}   
 \nn\\
&
+ \lambda_{S} (S^4+{\rm c.c.}) + \lambda'_{S} |S|^4
+{\lambda''_{S} |S|^2(S^2+{\rm c.c.})} 
+\lambda_{{\Phi S}}|\Phi|^2  (S^2+{\rm c.c.}) +\lambda'_{{\Phi S}}|\Phi|^2 |S|^2  
\nn\\&
+{\sum_{a,b=1}^N\left( \lambda_{{\Phi h_{ab}}}|\Phi|^2 h^+_a h^-_b 
 +\lambda_{{\Phi k_{ab}}}|\Phi|^2k^{++}_a k^{--}_b 
 +\lambda_{h_{ab} S}h^+_a h^-_b  (S^2+{\rm c.c.})  \right.
\left.+\lambda'_{h_{ab} S}h^+_a h^-_b  |S|^2
\right.}\nn\\&
{\left. +\lambda_{k_{ab}S}k^{++}_a k^{--}_b (S^2+{\rm c.c.}) 
+\lambda'_{k_{ab}S}k^{++}_a k^{--}_b |S|^2\right)},
\label{HP}
\end{align}
where we define $S=(S_R+iS_I)/\sqrt2$, $\tau_2$ is a second component of the Pauli matrix, the index $i(j)$ runs $1$-$3$, and $M_E$ can be diagonal without loss of the generality.
Notice here that $g$ should be symmetric matrix.
We work on the basis where all the coefficients are real and positive for our brevity. 
We parametrize  these scalar fields as 
\begin{align}
\Phi_{} &=\left[
\begin{array}{c}
\phi^+_{}\\
\phi^0_{}
\end{array}\right],\
\quad
\phi^0_{}=\frac1{\sqrt2}(v_{} + h_{} + ia_{}),
\label{component}
\end{align}
where $h$ is the SM-like Higgs, and $v$ is related to the Fermi constant $G_F$ by $v^2=1/(\sqrt{2}G_F)\approx(246$ GeV)$^2$.
The VEV of $\Phi_{}$ is derived from the tadpole condition $\partial \mathcal{V}/ \partial v=0$ such that
$v^2 \simeq \frac{m_{\Phi}^2}{\lambda_{\Phi}}$.
 After the electroweak symmetry breaking, we have massive gauge bosons $W_{}^\pm$ and $Z_{}$ by which the NG bosons $\phi^\pm$ and $a$ are absorbed respectively.
Each of scalar boson mass eigenvalue is straightforwardly obtained by the potential where we adopt these masses as a free parameters in our analysis below.
The charged lepton sector is the same as the SM lepton, that is, the charged lepton mass is given by the term of $y_\ell$ after the electroweak symmetry breaking.

The exotic vector-like SU(2)$_L$ doublet $L'$ is given as follow:
\begin{align}
L'_{L(R)}\equiv 
\left[
\begin{array}{c}
N\\
E^-
\end{array}\right]_{L(R)}.
\end{align}
{The mass term of $L'$ is given by $M_L (\bar E^-_L  E^-_R + \bar N_L  N_R)$ at the tree level. 
The mass difference of $E^-$ and $N$ is induced at the one-loop level via gauge interaction where the mass of $E^-$ becomes heavier by $O(300)$ MeV for $M_L \sim 1$ TeV~\cite{Cirelli:2005uq,Cirelli:2007xd}. 
For neutral component, we have Majorana fermions $N_L$ and $N_R^c$.}
Then its mass matrix in the basis of $[N_L,N^c_R]$ is given by 
\begin{align}
M_N=
\left[\begin{array}{cc}
\delta m & M_L \\
M_L &  0 \\
\end{array}\right],
\end{align}
where $\delta m$($<<M_L$) is induced at the Zee-Babu type of two-loop level ~\cite{zee-babu}, and we assume to be one generation case with positive real for our simple analysis. Here the loop function is the same as the Zee-Babu model.
$M_N$ is diagonalized by 2 $\times$ 2 unitary mixing matrix $V_N$ as
$V_N M_N V_N^T$ $\approx{\rm diag.}(M_L-\delta m/2, M_L + \delta m/2)\equiv{\rm diag.}(M_{N_1},M_{N_2}) $,  where we assume that $V_N$ is approximately a maximal mixing as follow:
\begin{align}
V_N\approx \frac{1}{\sqrt2}
\left[\begin{array}{cc}
i & -i \\
1 &1 \\
\end{array}\right],
\end{align}
then one finds our Majorana fields $\psi_1$ and $\psi_2$, by redefining  $\psi_1\equiv N_1+N^c_1$ and $\psi_2\equiv N^c_2+N_2$~\cite{Okada:2015nca}.
The lighter field $\psi_1$ can be a stable DM candidate and $\psi_2$ can be a meta-stable DM candidate , but the mass difference between them is tiny because $\delta m$ is generated at the two loop level. Thus  we  take account of the co-annihilation system including $\psi_1$, $\psi_2$ and $E^-$ to obtain the relic density.

\section{Phenomenology of the model}

In this section, we carry out phenomenological study of our model such as active neutrino mass, lepton flavor violation and dark matter, based on the set up discussed in the previous section.

\subsection{Active neutrino mass}

The active neutrino mass matrix  is generated at the three-loop level, but we compute the mass insertion method. Therefore the mass matrix is effectively induced at the one-loop level as follows: 
\begin{align}
(m_{\nu_{}})_{ab}&=
-\sum_{j=1}^{3} \frac{(f_R)_{aj} \delta m_j (f_R)_{bj} (m_{S_R}^2-m_{S_I}^2) M_{L_j}^2}{(4\pi)^2M_{\rm max}^4}
F_1(X_{L_j}, X_{S_R} , X_{S_I}),\\
F_1(X_{L_j}, X_{S_R} , X_{S_I})
&\equiv
\int dx dy dz \frac{x\delta(x+y+z-1) }{(x X_{L_j}+y X_{S_R} + z X_{S_I} )^2},
\label{eq:mD}
\end{align}
where 
we define
$X_{f}\equiv \left(\frac{m_{f} }{M_{\rm max}}\right)^2$ and $M_{\rm max}\equiv {\rm Max}[M_{L_j}, m_{S_R},m_{S_I} ]$.

Then $(m_{\nu_{}})_{ab}$ is diagonalized by the Maki-Nakagawa-Sakata mixing matrix $V_{\rm MNS}$ (MNS) as
\begin{align}
(m_{\nu_{}})_{ab} &=(V_{\rm MNS}^* D_\nu V_{\rm MNS}^\dag)_{ab},\quad D_\nu\equiv (m_{\nu_1},m_{\nu_2},m_{\nu_3}),
\\
V_{\rm MNS}&=
\left[\begin{array}{ccc} {c_{13}}c_{12} &c_{13}s_{12} & s_{13} e^{-i\delta}\\
 -c_{23}s_{12}-s_{23}s_{13}c_{12}e^{i\delta} & c_{23}c_{12}-s_{23}s_{13}s_{12}e^{i\delta} & s_{23}c_{13}\\
  s_{23}s_{12}-c_{23}s_{13}c_{12}e^{i\delta} & -s_{23}c_{12}-c_{23}s_{13}s_{12}e^{i\delta} & c_{23}c_{13}\\
  \end{array}
\right],
\end{align}
where we neglect the Majorana phase. 
The following neutrino oscillation data at 95\% confidence level~\cite{pdf} is given as
\begin{eqnarray}
&& 0.2911 \leq s_{12}^2 \leq 0.3161, \; 
 0.5262 \leq s_{23}^2 \leq 0.5485, \;
 0.0223 \leq s_{13}^2 \leq 0.0246,  
  \\
&& 
  \ |m_{\nu_3}^2- m_{\nu_2}^2| =(2.44\pm0.06) \times10^{-3} \ {\rm eV}^2,  \; 
  \ m_{\nu_2}^2- m_{\nu_1}^2 =(7.53\pm0.18) \times10^{-5} \ {\rm eV}^2, \nn
  \label{eq:neut-exp}
  \end{eqnarray}
where we assume one of three neutrino masses is zero with normal ordering in our analysis below.

To achieve the numerical analysis,  we apply the Casas-Ibarra parametrization~\cite{Casas:2001sr} to our form.
Thus the Yukawa coupling $f_R$ is rewritten as
\begin{eqnarray}
f_R = V^*_{\rm MNS} \sqrt{D_\nu} O \sqrt{R}, 
\label{yukawa}
\end{eqnarray}
where 
$O$, which  is an complex orthogonal matrix, and $R$, which is a diagonal matrix, are respectively formulated as
\begin{eqnarray}
O=
\left(
\begin{array}{ccc}
0 & 0 & 1 \\
\cos\alpha & \sin\alpha & 0 \\
-\sin\alpha & \cos\alpha & 0 \\
\end{array} 
\right),
\label{co}\quad \alpha\ {\rm is\ a \ complex\ parameter},
\end{eqnarray}
and
\begin{eqnarray}
R_{jj}=\frac{\delta m_j(m_{S_R}^2-m_{S_I}^2) M^2_{L_j}}{(4\pi)^2 M^4_{\rm max}} F_1(X_{L_j}, X_{S_R} , X_{S_I}).
\end{eqnarray}
Notice here that we assume the lightest neutrino mass is zero and the neutrino mass spectrum is normal hierarchy without  Majorana and Dirac phase in the numerical analysis below for brevity. The simplification of the mixing matrix $O$ is the direct consequence of its massless neutrino for the first generation.

\begin{table}[t]
\begin{tabular}{c|c|c} \hline
Process & $(i,j)$ & Experimental bounds ($90\%$ CL) \\ \hline
$\mu^{-} \to e^{-} \gamma$ & $(2,1)$ &
	$\text{Br}(\mu \to e\gamma) < 5.7 \times 10^{-13}$  \\
$\tau^{-} \to e^{-} \gamma$ & $(3,1)$ &
	$\text{Br}(\tau \to e\gamma) < 3.3 \times 10^{-8}$ \\
$\tau^{-} \to \mu^{-} \gamma$ & $(3,2)$ &
	$\text{Br}(\tau \to \mu\gamma) < 4.4 \times 10^{-8}$  \\ \hline
\end{tabular}
\caption{Summary of $\ell_i \to \ell_j \gamma$ process and the lower bound of experimental data~\cite{Adam:2013mnn}.}
\label{tab:Cif}
\end{table}

{\it  Lepton flavor violation processes}:
We  have the lepton flavor violations such as $\ell_i\to\ell_j\gamma$,~\footnote{
However there exist $\ell_i^-\to\ell_j^-\ell_k^+\ell^-_\ell$ processes at the one-loop level, we neglect them since they are subdominant processes~\cite{Toma:2013zsa}. }
and each  of flavor dependent process has to satisfy the current upper bound, as can be seen in Table~\ref{tab:Cif}. Notice here that
$\mu\to e\gamma$ process gives the most stringent upper bound.
Our branching fraction is given 
by
\begin{align}
\text{Br}(\ell_i\to\ell_j\gamma) &= 
\frac{3\alpha_{\rm em}  }{64 \pi  {\rm G_F}^2} \left| G_{ij}\right|^2, \\
G_{ij}&\equiv \sum_{k=1-3}(f_R)_{i,k}(f_R^\dag)_{kj} 
\left[ F_2\left(\frac{m^2_{S_R}}{M^2_{L_k}}\right) + F_2\left(\frac{m^2_{S_I}}{M^2_{L_k}}\right) \right],\\
F_2(x)&\equiv \frac{1-6x+3x^2+2x^3-6x^2 \ln x}{6(1-x)^4},
\end{align}
where ${\rm G_F}\approx 1.1\times 10^{-5}[{\rm GeV}^{-2}]$ is Fermi constant, and $\alpha_{\rm em}\approx1/137$ is the fine-structure constant.

{\it Numerical results}:
Now we discuss the numerical solutions to satisfy the neutrino oscillation data and the LFV constraints.
Here we randomly select the values of 
these parameters within the following ranges:  
\begin{eqnarray}
&M_{L_{1}} =\left(100 \ {\rm GeV}, 2000\ {\rm GeV} \right), 
M_{L_{2,3}} =\left(M_{L_{1}} , 2000\ {\rm GeV} \right), 
m_{S_R} =\left(M_{L_{1}} , 2000 \ {\rm GeV}\right), \nn\\ 
& m_{S_I}=\left(M_{L_1}, 2000 \ {\rm GeV}\right),
 \delta m = \left(200 \ {\rm keV}, 400 \ {\rm keV}\right),
\end{eqnarray}
where all the elements of $f_R$ is taken to be $f_R < \sqrt{4\pi}$ as the perturbative limit, and
$10^5$ sampling points are our examinations to search for our allowed parameters.
 We find that 386 solutions over all the range that we take, and all the typical absolute order of $f_R$ is ${\cal O}$(0.01).

\subsection{ Estimation of the scale $\delta m$}
 In this section we estimate the scale of $\delta m$ that is generated at the two-loop level through the Zee-Babu like diagram, where we assume that all the off-diagonal elements for related Yukawa couplings ($f_L$ and $g$) are zero. Thus we can  evade any kinds of LFV processes, even when we take rather larger values of diagonal Yukawa couplings. Then the mass formula is given by
\begin{align}
\delta m&\approx \frac{8 {N^3}\mu f_L^2 g m_\tau^2}{(4\pi)^4 m_{k^{\pm\pm}}^2}F_3(X),\\
F_3(X)&\equiv \int_0^1 dx \int_0^1 d\alpha \int _0^{1-y} dy  \int _0^{1-\alpha} d\beta  
[(y^2-y)\alpha X - \beta (x X+y)]^{-1},
\end{align}
where $X\equiv (m_{h^\pm}/m_{k^{\pm\pm}})^2$, $m_\tau(=1.776$ GeV)  is the tau lepton mass, and neglect the masses of the charged leptons in the loop function $F_3(X)$. 
{We assume $m_{h_1}=m_{h_2}=\cdots m_{h}, m_{k_1}=m_{k_2}=\cdots m_{k}, f_{L}^1=f_{L}^2=\cdots f_L$
and $g^1=g^2=\cdots g .$}
Then it simplifies $F_3(X)\approx 1/X$. 
{\it The important constraint comes from the inelastic direct detection searches via $Z$ boson portal that
suggests that  ${\cal O}(200\ {\rm keV})\lesssim\delta m$.}
Also considering the correction bound for the charged boson masses, the trilinear coupling $\mu$ can roughly be restricted as $\mu\lesssim (4\pi) {\rm Min}[m_h^{\pm},m_k^{\pm\pm}]$~\cite{Herrero-Garcia:2014hfa}. Here we assume $m_k^{\pm\pm}<< m_h^{\pm}$ for our convenience and adopt the upper limit of the trilinear coupling $\mu$ in the following analysis.
{Moreover, we hereafter fix the mass of $k^{\pm\pm}$ to be $m_k^{\pm\pm} \approx 700$ GeV which is sufficiently larger than the lowest bound by doubly charged scalar search at the LHC~\cite{cms-exp}. }
Applying these requirements to the mass formula, we obtain the following relation
\begin{align}
\frac{0.05 X m_{k^{\pm \pm}} [{\rm GeV}]}{m_\tau^2}\lesssim (NY)^3\to 
\left[\frac{0.05 X m_{k^{\pm \pm}} [{\rm GeV}]}{m_\tau^2}\right]^{1/3}\lesssim NY,\label{eq:cond1}
\end{align} 
where we assume $Y\equiv f_L=g$ for brevity.
In this case we have to be careful of the fact that the $N$ negatively enlarges the $(g-2)_\mu$, although the LFVs are still zero as far as the diagonal $Y(=f_L=g)$.
We thus consider the upper limit for the absolute value of $(g-2)_\mu$ which is given by
\begin{align} 
 |\Delta a_\mu|\approx \left|-\frac{|NY|^2}{3(4\pi)^2}\frac{m_\mu^2}{m^2_{k^{\pm\pm}}}\right| \lesssim 40\times 10^{-10},
 \end{align}
 where the last inequality represents the absolute upper bound from  the experimental result in ref.~\cite{discrepancy2}.
Combining with Eq.~(\ref{eq:cond1}) and $m_k^{\pm\pm}\approx 700$ GeV, we obtain the following upper mass bound  for the singly charged bosons $h^\pm$
\begin{align}
m_{h^\pm}\lesssim 6\ {\rm TeV}. \label{eq:cond2}
\end{align}
The upper bound plays an role in determining the lower mass bound of $h^\pm$ in  analyzing the DM relic density, as will be discussed below.

\subsection{Dark Matter}
We consider fermionic DM candidates $X(\equiv \psi_{1})$, which is assumed to be the lightest and stable particle, and  $\psi_{2}$, which is assumed to be the meta-stable particle. Here we analyze the one flavor case for simplicity.
Since the mass difference between $\psi_1$ and $\psi_2$ is tiny due to generating the two-loop level, we have to consider the measured relic density on the co-annihilation system. 
Furthermore, the mass of $E^-$ is also close to that of DM.
Thus we assume to be $M_X\approx M_{\psi_2}\approx M_E$ throughout the DM analysis.

\begin{figure}[t]\begin{center}
\includegraphics[width=0.80\columnwidth]{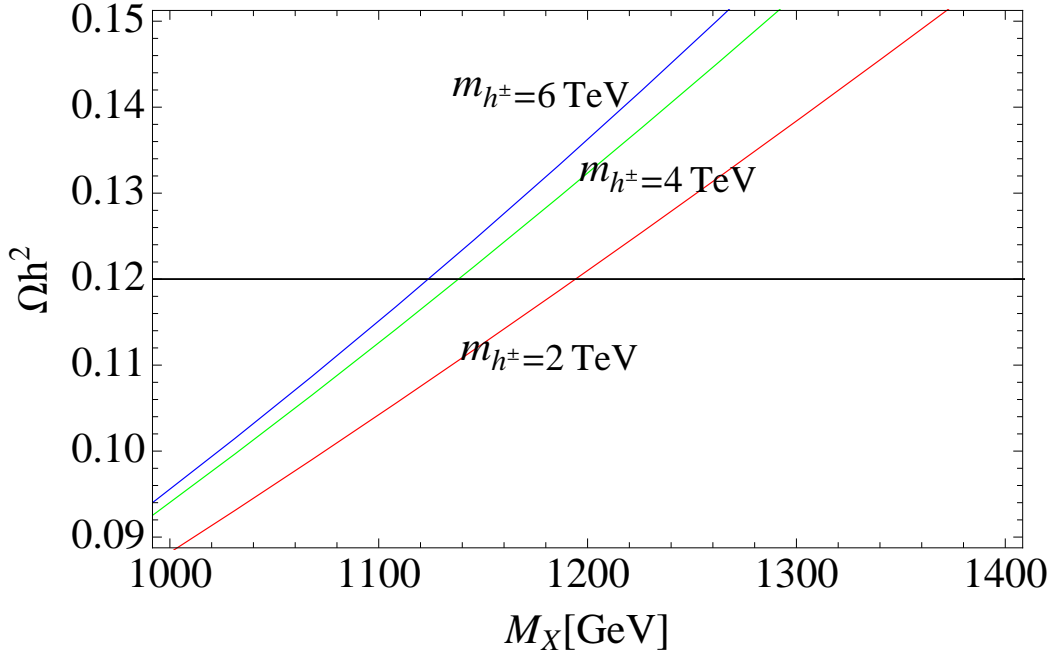}
   \caption{The relic density in terms of  the DM mass. Each of the blue, green, and red line corresponds to $m_{h^\pm}=(6,4,2)$ TeV. The black horizontal is the observed relic density; $\Omega h^2\approx0.12$.}   \label{relic-dm}
\end{center}\end{figure}

{\it Relic density}:
{The relic density of DM is obtained by calculating annihilation and coannihilation cross sections where 
our dominant contribution comes from the kinetic term of $L'$ and the term with $f_L$. }
For gauge interactions, the complete analysis is provided in ref.~\cite{Cirelli:2007xd}.
In addition to gauge interactions,  the (co-)annihilation cross section from the Yukawa interactions is given by
\begin{align}
\sigma v_{\rm rel}\approx \frac{|NY|^4 r^2(1-2r+2r^2)}{48\pi M_X^2} v_{\rm rel}^2,
\end{align}
where $r\equiv M_X^2/(m^2_{h^\pm}+M_X^2)$. Notice here that we eliminate the $NY$ dependence by substituting Eq.~(\ref{eq:cond1}).
Adding the above p-wave to the effective whole p-wave, we estimate the relic density in terms of the DM mass as can be seen in Fig.~\ref{relic-dm}, applying the approximated formulae in refs.~\cite{Okada:2015nca, Griest:1990kh}.
In this figure, each of the colored line; (blue, green, red), represents $m_{h^\pm}=(6,4,2)$ TeV, and the black horizontal is the observed relic density; $\Omega h^2\approx0.12$~\cite{Ade:2013zuv}. Notice here that $m_{h^\pm}=6$ TeV comes from the upper constraint for the $(g-2)_\mu$ discussed in Eq.~(\ref{eq:cond2}).

In case where $m_{h^\pm}=6$ TeV,  
the Yukawa contribution is negligible due to the suppression of the heavy mass of $h^\pm$, and
the allowed DM mass is about $1.12$ TeV, which is consistent of the result to the ref~\cite{Cirelli:2007xd}.~\footnote{In this reference, they consider the gauge interactions only.}
It suggests that the lower mass of the singly charged boson is $1.12 $ TeV to kinematically forbid the too rapid two body decay $X\to h^+\ell$ through the term $f_L$.  Therefore, we obtain the following 
relation 
\begin{align}
1.12\ {\rm TeV}\lesssim m_{h^\pm}\lesssim 6\ {\rm TeV}.
\end{align}
{In case where $m_{h^\pm}=(4,2)$ TeV,  on the other hand, the contribution from Yukawa interactions becomes to be relevant in relic density estimation. Thus the value of relic density decrease at the same DM mass since the (co-)annihilation cross sections become larger.}
As a result, we obtain the allowed DM mass at $(1.14, 1.2)$ TeV, respectively.

{\it Photon excess}:
It might be worth mentioning the photon excess such that the emitting photon energy $E_\gamma$ can be replaced by the mass difference ($\delta m$)  between $\psi_1$ and $\psi_2$.
Since our model has the lower bound; ${\cal O}(200)\ {\rm keV}\lesssim E_\gamma$ to evade the constraint from direct detection searches with spin independent inelastic scattering process via $Z$ boson portal, the X-ray line at the 3.55 keV energy cannot be explained.
However, we still expect that the other experiments might detect our energy scale near future.  As one of the examples, we focus on  the 511 keV photon excess by INTEGRAL~\cite{Knodlseder:2003sv} and briefly discuss. The INTEGRAL experiment provides the gamma ray line flux at around 511 keV as
\begin{align}
\Phi_\gamma(511{\rm keV})=(1.05\pm 0.06)\times 10^{-3}\ {\rm photon\ cm}^{-2}{\rm s}^{-1}.
  \end{align}
On the other hand, our gamma ray flux is found as~\cite{Khalil:2008kp}
\begin{align}
\Phi_\gamma(511{\rm keV})\approx\frac{10^{27}{\rm s}  } {2}
\left[\frac{\Gamma(X_2\to X\gamma)}{M_X  ({\rm MeV}^{-1})}\right]\times 10^{-3} \ {\rm photon\ cm}^{-2}{\rm s}^{-1},
\end{align}
where 
\begin{align}
& \frac{\Gamma(X_2\to X\gamma)}{M_X}\approx
\frac{\alpha_{\rm em} \delta m^3 M_X}{64\pi^4 m_{h^\pm}^4}
\left|
\sum_{k}^{1-3}{\rm Im} [ (f_R)_{k1} (f_R)^*_{k2}]
G\left(\frac{M_X^2}{m_{h^\pm}^2} 
\right)\right|^2,\\
& G(X 
)
\equiv \int_0^1\frac{dx x(x-1)}{x^2 X-(1+X
)x+1}.
\label{eq:x-ray}
\end{align}
Comparing with our formula and the experimental result, we obtain the following constraint for the Yukawa couplings $f_R$
\begin{align}
&1.9\times 10^{-10}\lesssim \left|\sum_{k}^{1-3}{\rm Im} [ (f_R)_{k1} (f_R)^*_{k2}]\right|\lesssim 2.0\times 10^{-10} \ {\rm for}\ {m_{h^\pm}=6}\ {\rm TeV}(M_X=1.12\ {\rm TeV}),\\
& 8.2\times 10^{-11}\lesssim \left|\sum_{k}^{1-3}{\rm Im} [ (f_R)_{k1} (f_R)^*_{k2}]\right|\lesssim 8.7\times 10^{-10} \ {\rm for}\ {m_{h^\pm}=4}\ {\rm TeV}(M_X=1.14\ {\rm TeV}),\\
& 1.6\times 10^{-11}\lesssim \left|\sum_{k}^{1-3}{\rm Im} [ (f_R)_{k1} (f_R)^*_{k2}]\right|\lesssim 1.7\times 10^{-11} \ {\rm for}\ {m_{h^\pm}=2}\ {\rm TeV}(M_X=1.2\ {\rm TeV}),
\end{align}
where we have used each the solution for $(m_{h^\pm},M_X)$ in analyzing  the relic density.
It suggests that the Yukawa CP phases are very small, and this can easily be realized, satisfying the neutrino oscillation data.

\section{Conclusions and Discussions}
We have studied radiatively {induced} neutrino mass model, in which we have used the mass insertion method to the neutrino sector.
As a result, neutrino mass is effectively generated at the one-loop level with the degenerated neutral fermions masses which obtain mass splitting at the two loop level.
Considering the constraints of  LFV processes, we have numerically analyzed  the allowed region search to satisfy the LFVs and the neutrino oscillation data.
Then we have discussed the DM phenomenologies, in which we have firstly investigated the constraint of inelastic scattering direct detection searches through the SM neutral vector boson portal. The experiment suggests that the mass difference between the degenerated masses should be larger than the order ${\cal O}$
(200 keV). We have found the solution to evade this constraint by introducing the multiple charged bosons with the same quantum number as these fields that we originally be introduced. This modification is in favor of the recent experimental result of the diphoton excess at the 750 GeV reported by CMS~\cite{CMS:2015dxe} and ATLAS~\cite{ATLAS-CONF-2015-081} where multiple charged scalars  enhance diphoton decay branching ratio of 750 GeV neutral scalar interacting with these charged scalars~\cite{Kanemura:2015bli,Nomura:2016fzs,Yu:2016lof,Ding:2016ldt,Nomura:2016seu,Okada:2016rav,Nomura:2016rjf,Ko:2016sxg,Hernandez:2015hrt,Arbelaez:2016mhg,Chao:2015nac}. However it simultaneously induces the negative enhancement of the muon anomalous magnetic moment. Considering the upper absolute value
$|\Delta a_\mu|  \lesssim40\times 10^{-10}$, we have obtained the upper bound for the singly charged boson mass; $m_{h^\pm}\lesssim$6 TeV. 
Applying this upper bound, we have analyzed the DM relic density with co-annihilation system and found that the dominant annihilation mode comes from the kinetic term of $L'$ in case of $m_{h^\pm}=$ 6 TeV. But if  the mass of ${h^\pm}$ decreases, the value of relic density also decreases at the same DM mass, therefore the allowed DM mass increases in fig.~\ref{relic-dm}. 
Then we have briefly discussed the possibility to detect  photon emission through indirect detection searches larger than  our threshold energy $\simeq {\cal O}$
(200 keV). Here we have focused on the INTEGRAL experiment at 511 keV photon energy, and we have found that the imaginary part of  Yukawa coupling $f_R$ is constrained as $\left|\sum_{k}^{1-3}{\rm Im} [ (f_R)_{k1} (f_R)^*_{k2}]\right| \approx {\cal O}(10^{-10}-10^{-11})$, depending on our three bench mark sets of $(m_{h^\pm},M_X)$ obtained by the analysis of relic density.

It might be worth mentioning the muon anomalous magnetic moment in our model.
We have new sources of positive contribution from $f_R$ and negative contribution from $f_L$,
but the positive contribution is constrained by the LFVs, therefore $f_R$ is typically the order ${\cal O}(0.01)$ as  discussed before.
Hence our total $(g-2)_\mu$ tends to be negative value, which is against the experimental result. Thus we have used this result as the upper bound to restrict the charged boson. If future experiments could provide  a negative contribution, it could still be verifiable.

{The extra charged particles could be produced at the LHC since the masses of these particles are around O(1) TeV in our model. 
Thus we can test our model by future experiments searching for exotic charged particles. The detailed analysis is beyond the scope of our study and will be given in future work.}

\section*{Acknowledgments}
\vspace{0.5cm}
Authors thank to Dr. Kei Yagyu for fruitful discussions.
H.O. expresses his sincere gratitude toward all the KIAS members, Korean cordial persons, foods, culture, weather, and all the other things.
This work was supported by the Korea Neutrino Research Center which is established by the National Research Foundation of Korea(NRF) grant funded by the Korea government(MSIP) (No. 2009-0083526) (Y.O.).


\end{document}